\documentclass[conference,compsoc,onecolumn]{IEEEtran}
\ifCLASSOPTIONcompsoc
  \usepackage[nocompress]{cite}
\else
  \usepackage{cite}
\fi

\ifCLASSINFOpdf
\else
\fi
\usepackage{amsmath}

\usepackage{array}
\ifCLASSOPTIONcompsoc
\usepackage[caption=false,font=footnotesize,labelfont=sf,textfont=sf]{subfig}
\else
\usepackage[caption=false,font=footnotesize]{subfig}
\fi
\usepackage{url}
\usepackage{graphicx}

\usepackage{algorithm,algpseudocode}
\algrenewcommand\algorithmicrequire{\textbf{Input:}}
\algrenewcommand\algorithmicensure{\textbf{Output:}}
\usepackage{amsmath,amsfonts,amssymb,amsthm}
\usepackage{xcolor}
\definecolor{ocre}{RGB}{0,173,239}

\usepackage{tikz}
\usepackage{pgfplots}
\pgfplotsset{compat=1.18}

\usetikzlibrary{arrows,matrix,positioning,fit}
\usetikzlibrary{patterns,decorations.pathreplacing}
\tikzset{%
  highlight/.style={rectangle,rounded corners,fill=ocre!50,draw=ocre!80,
     fill opacity=0.2,inner sep=0pt,text opacity=1}
}

\usepackage{graphicx} 
\usepackage{float}
\usepackage[all]{xy}
\usepackage{tikz-cd}
\usepackage[shortlabels]{enumitem}
\usepackage{dsfont}
\usepackage{tikz}
\usepackage{pbox}
\usetikzlibrary{backgrounds}
\pgfdeclarelayer{background2}
\pgfdeclarelayer{background}
\pgfdeclarelayer{foreground}
\pgfdeclarelayer{foreground2}
\pgfsetlayers{background2,background,main,foreground,foreground2}

\makeatletter
\renewcommand*\env@matrix[1][*\c@MaxMatrixCols c]{%
  \hskip -\arraycolsep
  \let\@ifnextchar\new@ifnextchar
  \array{#1}}
\makeatother

\usepackage[edges]{forest}
\usepackage{array}
\newcolumntype{C}[1]{>{\centering\arraybackslash}p{#1}}
\newcolumntype{L}[1]{>{\raggedright\arraybackslash}p{#1}}
\usetikzlibrary{arrows.meta,shadows}
\usetikzlibrary{trees,positioning,shapes,shadows,arrows}

\usepackage[hypertexnames=false]{hyperref}
\hypersetup{pdfborder=0 0 0}

\algnewcommand{\LineComment}[1]{\State \(\triangleright\) #1}

\usepackage{tablefootnote} % for table footnotes

\usepackage{lastpage} % for page out of
\usepackage{fancyhdr} % for page out of
\newlength{\gpgpuElemSep}
\newlength{\gpgpuElemSize}
\usepackage{listings}
\usepackage{caption}
\makeatletter
\let\old@lstKV@SwitchCases\lstKV@SwitchCases
\def\lstKV@SwitchCases#1#2#3{}
\makeatother
\usepackage{lstlinebgrd}
\makeatletter
\let\lstKV@SwitchCases\old@lstKV@SwitchCases

\lst@Key{numbers}{none}{%
    \def\lst@PlaceNumber{\lst@linebgrd}%
    \lstKV@SwitchCases{#1}%
    {none:\\%
     left:\def\lst@PlaceNumber{\llap{\normalfont
                \lst@numberstyle{\thelstnumber}\kern\lst@numbersep}\lst@linebgrd}\\%
     right:\def\lst@PlaceNumber{\rlap{\normalfont
                \kern\linewidth \kern\lst@numbersep
                \lst@numberstyle{\thelstnumber}}\lst@linebgrd}%
    }{\PackageError{Listings}{Numbers #1 unknown}\@ehc}}
\makeatother

\def\FunctionD(#1){2*(#1)-1}%
\def\FunctionM(#1,#2){((3*(#1)^2)*10^9)/(#2)}
\newcommand{\PreserveBackslash}[1]{\let\temp=\\#1\let\\=\temp}
\newcommand\hlightcolor[1]{\tikz[overlay, remember picture,baseline=-\the\dimexpr\fontdimen22\textfont2\relax]\node[rectangle,fill=magenta!50,rounded corners,fill opacity=0.2,draw=magenta!80,text opacity=1] {$#1$};} 
\usepackage{soul} %strikethrough font
\setstcolor{blue}
\usepackage{accents}

\definecolor{myblue4} {RGB}{66,129,196}
\tikzstyle{myline} = [very thick, draw=myblue4, fill=myblue4, shape=rectangle, inner sep=10pt, inner ysep=20pt]
\tikzstyle{mybox} = [very thick, draw=myblue4, fill=white, shape=rectangle, inner sep=10pt, inner ysep=20pt]

\tikzstyle{mybox1} = [very thick, draw=white, fill=white, shape=rectangle, inner sep=10pt, inner ysep=20pt]

\usepackage{adjustbox}
\usepackage{fourier-orns}
\usepackage{multirow}
\usepackage{tikz}
\usepackage{pgfplotstable}
\usepackage{ifthen}

\DeclareCaptionFormat{algor}{%
\hrulefill\par\offinterlineskip\vskip1pt%
\textbf{#1#2}#3\offinterlineskip\hrulefill}
\DeclareCaptionStyle{algori}{singlelinecheck=off,format=algor,labelsep=space}
\usepackage{bm}
\usepackage{setspace}

\usepackage{multicol}
\usepackage{pdflscape}

\usepackage{longtable}
\usepackage{stackengine}

\usepackage{subcaption}
\selectcolormodel{gray}
\pgfplotscreateplotcyclelist{my_black_white}{%
solid, every mark/.append style={solid, fill=gray}, mark=*\\%
solid, every mark/.append style={solid, fill=gray}, mark=square*\\%
solid, every mark/.append style={solid, fill=gray}, mark=star\\%
solid, every mark/.append style={solid, fill=gray}, mark=diamond*\\%
solid, every mark/.append style={solid, fill=gray}, mark=otimes\\%
solid, every mark/.append style={solid, fill=gray}, mark=Mercedes star\\%
solid, every mark/.append style={solid, fill=gray}, mark=triangle*\\%
solid, every mark/.append style={solid, fill=gray}, mark=oplus\\%%%%%
densely dotted, every mark/.append style={solid, fill=gray}, mark=*\\%
densely dotted, every mark/.append style={solid, fill=gray}, mark=square*\\%
densely dotted, every mark/.append style={solid, fill=gray}, mark=star\\%
densely dotted, every mark/.append style={solid, fill=gray}, mark=diamond*\\%
densely dotted, every mark/.append style={solid, fill=gray}, mark=otimes\\%
densely dotted, every mark/.append style={solid, fill=gray}, mark=Mercedes star\\%
densely dotted, every mark/.append style={solid, fill=gray}, mark=triangle*\\%
densely dotted, every mark/.append style={solid, fill=gray}, mark=oplus\\%%%%%
dashed, every mark/.append style={solid, fill=gray}, mark=*\\%
dashed, every mark/.append style={solid, fill=gray}, mark=square*\\%
dashed, every mark/.append style={solid, fill=gray}, mark=star\\%
dashed, every mark/.append style={solid, fill=gray}, mark=diamond*\\%
dashed, every mark/.append style={solid, fill=gray}, mark=otimes\\%
dashed, every mark/.append style={solid, fill=gray}, mark=Mercedes star\\%
dashed, every mark/.append style={solid, fill=gray}, mark=triangle*\\%
dashed, every mark/.append style={solid, fill=gray}, mark=oplus\\%%%%%
dashdotted, every mark/.append style={solid, fill=gray}, mark=o\\%
dashdotted, every mark/.append style={solid, fill=white}, mark=square*\\%
}
\begin{document}
%\thispagestyle{empty}
%Number of pages: \pageref{LastPage}
%\tableofcontents
%\listoffigures
%\listoftables
%\listofalgorithms
%\lstlistoflistings
%\newpage
\pagenumbering{arabic}
\pagestyle{plain}

\title{ML-Based Optimum Number of CUDA Streams\\ for the GPU Implementation of the Tridiagonal Partition Method}

% \author{\IEEEauthorblockN{Milena Veneva}
% \IEEEauthorblockA{RIKEN Center for Computational Science, R-CCS, 7-1-26 Minatojima-minami-machi, Chuo-ku, \\Kobe, Hyogo 650-0047, Japan,\\
% Email: milena.p.veneva@gmail.com}
% \and
% \IEEEauthorblockN{Toshiyuki Imamura}
% \IEEEauthorblockA{RIKEN Center for Computational Science, R-CCS, 7-1-26 Minatojima-minami-machi, Chuo-ku, \\Kobe, Hyogo 650-0047, Japan,\\
% Email: imamura.toshiyuki@riken.jp}}
\author{
\IEEEauthorblockN{Milena Veneva\IEEEauthorrefmark{1}, \emph{milena.p.veneva@gmail.com},\\Toshiyuki Imamura\IEEEauthorrefmark{1}, \emph{imamura.toshiyuki@riken.jp}}
%\vspace{0.05in}
\IEEEauthorblockA{\IEEEauthorrefmark{1}RIKEN Center for Computational Science, R-CCS, 7-1-26 Minatojima-minami-machi, Chuo-ku, \\Kobe, Hyogo 650-0047, Japan}
%\vspace{0.05in}
} 

% make the title area
\maketitle
\thispagestyle{plain}

\section*{Abstract}
This paper presents a heuristic for finding the optimum number of CUDA streams by using tools common to the modern AI-oriented approaches and applied to
the parallel partition algorithm. 
A time complexity model for the GPU realization of the partition method is built. Further, a refined time complexity model for the partition algorithm being executed on multiple CUDA streams is formulated. 
Computational experiments for different SLAE sizes are conducted, and the optimum number of CUDA streams for each of them is found empirically.
Based on the collected data a model for the sum of the times for the non-dominant GPU operations (that take part in the stream overlap) is formulated using regression analysis. 
A fitting non-linear model for the overhead time connected with the creation of CUDA streams is created.
Statistical analysis is done for all the built models.
An algorithm for finding the optimum number of CUDA streams is formulated. Using this algorithm, together with the two models mentioned above, predictions for the optimum number of CUDA streams are made. Comparing the predicted values with the actual data, the algorithm is deemed to be acceptably good. 

%%%%%%%%%%%%%%%%%%%%%%%%%%%%%%%%%%%%%%%%%%%%%%%%%%%%%%%%%%%%%%%%%%%%%%%
\section{Introduction} 
The parallel partition algorithm for solving systems of linear algebraic equations~(SLAEs) 
%with tridiagonal coefficient matrices 
suggested in~\cite{Austin:2004} is an efficient numerical approach for solving SLAEs with tridiagonal
coefficient matrices, splitting the matrix into sub-matrices and then solving
smaller SLAEs in parallel. The algorithm was initially intended for a large number
of processors and was implemented with the help of the MPI technology in~\cite{Austin:2004}. 

The development of HPC applications involves two major steps: developing correct code, and improving the code for performance. We present one of the optimizations made to our CUDA~\cite{cuda} implementation, namely building a heuristic for finding the optimum number of CUDA streams by using tools common to the modern AI-oriented approaches. 

CUDA devices contain engines for various tasks, e.\,g.,\,memory copy and kernel execution. If memory transfers and kernels' execution are dispatched into different streams, these operations can be overlapped.
The copy-compute overlap allows us to hide, partially or entirely, the memory
transfers behind computations (or the other way around), and thus shorten the total
time for the program. However, after
a certain amount of streams, the performance would stop
increasing, because the overhead of creating the additional
streams would be bigger than the actual execution time
boost. Thus, having a heuristic which predicts the optimum number of CUDA streams for each SLAE size would allow us to gain uttermost of the GPU performance.

Most of the known heuristics~(for instance, \cite{Werkhoven:2014}) depend on the times
for the memory transfers and the computational kernels
being executed without copy-compute overlap 
%which requires additional runs every time. 
which demands further trials constantly.
This is impractical because it increases the computational time on the respective supercomputer or cluster. Therefore, we suggest a different approach.

%%%%%%%%%%%%%%%%%%%%%%%%%%%%%%%%%%%%%%%%%%%%%%%%%%%%%%%%%%%%%%%%%%%%%%%
\section{Building a heuristic for the optimum number of CUDA streams}

The heuristic processes of optimization in solving SLAEs on GPUs are summarized as
%The computational experiments were held on the basis 
of NVIDIA GPU RTX 2080 Ti~\cite{2080a,2080b}, for SLAE sizes $10^{i}$, $2.5\times10^{i}$, $4\times10^{i}$, $5\times10^{i}$, $7.5\times10^{i}$, and $8\times10^{i}$, $i=3,4,\dots,7$, sub-system size equal to $10$; 256 CUDA threads per block; FP64 precision. 

\subsection{Hardware working queues and CUDA streams.}
The Hyper-Q technology allows
for multiple host threads to schedule work on the GPU
simultaneously by using multiple host-device hardware connections. In case Hyper-Q is not supported by the device, the
concurrency that can be achieved by copy-compute overlap
is limited. If the number of created streams is bigger than
the number of hardware working queues, multiple streams use one and the same queue. As the maximum number of hardware working queues for devices that support Hyper-Q is 32, using more than 32 streams is going to lead to serialization. Hence, we are considering only number of streams that are powers of 2, up to 32. 

\subsection{Time complexity model.}
\begin{figure}[htb]
\centering
\subfloat[]{{\fbox{\includegraphics[width=0.5\textwidth]{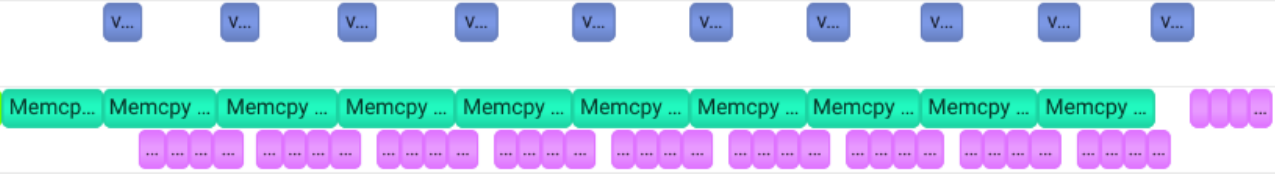}}}}\qquad
\subfloat[]{{\fbox{\includegraphics[width=0.5\textwidth]{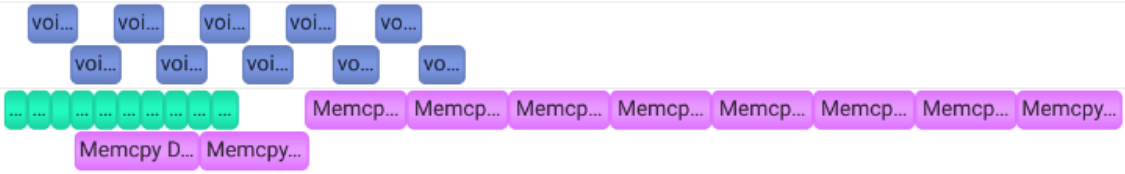}}}}%
\caption{Nsight Systems profiles of the kernel responsible for Stage 1 (a), and the kernel responsible for Stage 3 (b); SLAE size $10^{8}$, sub-system size 10, 10 CUDA streams. Cyan/pink boxes denote H2D/D2H memory transfers; blue -- kernels.}
\label{fig:nsys_k1_k_3}
\end{figure}

The time complexity model for the partition method (consisting of two stages on the GPU -- Stage 1, and 3, and one on the CPU -- Stage 2) is:
\begin{equation}
T_{non\_str} = \left(T_{1}^{H2D} + T_{1}^{COMP} + T_{1}^{D2H}\right) + T_{2}^{\mathrm{COMP}} +\left(T_{3}^{H2D} + T_{3}^{COMP} + T_{3}^{D2H}\right),
\end{equation}
where $T_{i}^{*}$ is the memory transfer/kernel time of Stage~$i, i=1,2,3$. 
Applying copy-compute overlap, and recalling that the algorithm is memory-bound, then the time for the computational kernels is going to be effectively hidden behind the memory transfers. Taking into account algorithm's nature, it is clear that $T_{1}^{H2D} > T_{1}^{D2H}$, and $T_{3}^{H2D} < T_{3}^{D2H}$. Up to an SLAE size $\leq10^{5}$, $T_{1}^{COMP} > T_{1}^{D2H}$, and $T_{3}^{COMP}> T_{3}^{D2H}$.
Thus, the refined model is:
\begin{equation}
\label{eq:time_comp}
T_{\textrm{str}} = T_{1}^{H2D} + \frac{T_{1}^{COMP} + T_{1}^{D2H} + T_{3}^{H2D}+T_{3}^{COMP}}{\textrm{num\_str}} + T_{2}^{\mathrm{COMP}} + T_{3}^{D2H} + T{\mathrm{\_overhead}},
\end{equation}
where $T\mathrm{\_{overhead}}$ is the overhead from the creation of CUDA streams. This is a lower bound estimation for $T_{\textrm{str}}$ ($T_{1}^{H2D}$ is not always big enough to hide both $T_{1}^{COMP}$, and $T_{1}^{D2H}$). Also, the GPU experiences idle times for SLAE sizes that do not saturate it. The same is true for Stage 3. This model is exact for big SLAE sizes though.
The model is proven by the NVIDIA Nsight systems~\cite{nsys} profiles in Figure~\ref{fig:nsys_k1_k_3}.

\subsection{Overview of existing heuristics.}
The overhead of forking CUDA streams in~\cite{Gómez-Luna:2011}
is calculated as the product of the number of streams, and time for creating one stream $\tau$ which should be computed for each GPU (it was found experimentally that $\tau = 0.004448$ ms for NVIDIA RTX 2080 Ti).
Thus, we can find the optimum number of streams by finding the first derivative of the time complexity model with regard to the number of streams. As a result,
the optimum number of CUDA streams depends only on $T_{1}^{COMP}$, $T_{1}^{D2H}$, $T_{3}^{H2D}$, $T_{3}^{COMP}$ (all measured when no streams are used), and $\tau$.
Applying this approach to SLAE sizes $4\times10^{i},~i~=~3,4,\dots,7$, it becomes clear that this model does not work well for the partition method, because it predicts a much higher number of streams than needed (for SLAE size $4\times10^{3}$: 8 vs.\,1; for SLAE size $4\times10^{7}$: 140 vs.\,32, etc.). The results can be seen in Table~\ref{tab:num_str_opt_investigation}.
Figure~\ref{fig:t_overhead_all_data} also proves that this approximation does not satisfy the empirical data.
We should consider a different model.

\begin{table}[htb]
%{\footnotesize
\centering
\caption{Times for the GPU operations of the partition method which take part in the stream overlap (given in [ms]) for SLAE sizes $4~\times~10^{i},\,i~=~3,4,\dots,7$, and comparison between the optimum number of CUDA streams according to \cite{Gómez-Luna:2011} (column {\it optimum streams} \cite{Gómez-Luna:2011}), and according to the computational experiments (column {\it actual optimum streams}).}
\label{tab:num_str_opt_investigation}
\begin{tabular}{|r|r|r|r|r|r|r|r|}
\hline
% size & $T_1^{COMP}$ & $T_1^{D2H}$ & $T_3^{H2D}$ & $T_3^{COMP}$ & sum & \cite{Gómez-Luna:2011} & actual \\
% & & & & & & optimum & optimum \\
% & & & & & & num\_str & num\_str \\\hline
size & $T_1^{COMP}$ & $T_1^{D2H}$ & $T_3^{H2D}$ & $T_3^{COMP}$ & sum & optimum & actual \\
& & & & & & streams & optimum \\
& & & & & & \cite{Gómez-Luna:2011} & streams \\\hline
$4\times10^{3}$ & 0.221312 & 0.014848 & 0.006592 & 0.030688 & 0.273440 & 7.8 & 1\\\hline 
$4\times10^{4}$ & 0.216544 & 0.057312 & 0.015456 & 0.038112 & 0.327424 & 8.6 & 1\\\hline 
$4\times10^{5}$ & 0.393184 & 0.402944 & 0.102784 & 0.205408 & 1.104320 & 15.8 & 4\\\hline 
$4\times10^{6}$ & 1.993980 & 3.897410 & 0.975392 & 2.130500 & 8.997282 & 45.0 & 32\\\hline 
$4\times10^{7}$ & 17.451500 & 38.836800 & 9.606720 & 20.981600 & 86.876620 & 139.8 & 32\\\hline 
\end{tabular}%}
\end{table}

\subsection{Building mathematical models for $\textrm{sum}$, and $T\mathrm{\_{overhead}}$.}
\begin{figure}[htb]
\centering
\includegraphics[trim={0cm 15cm 0 0}]{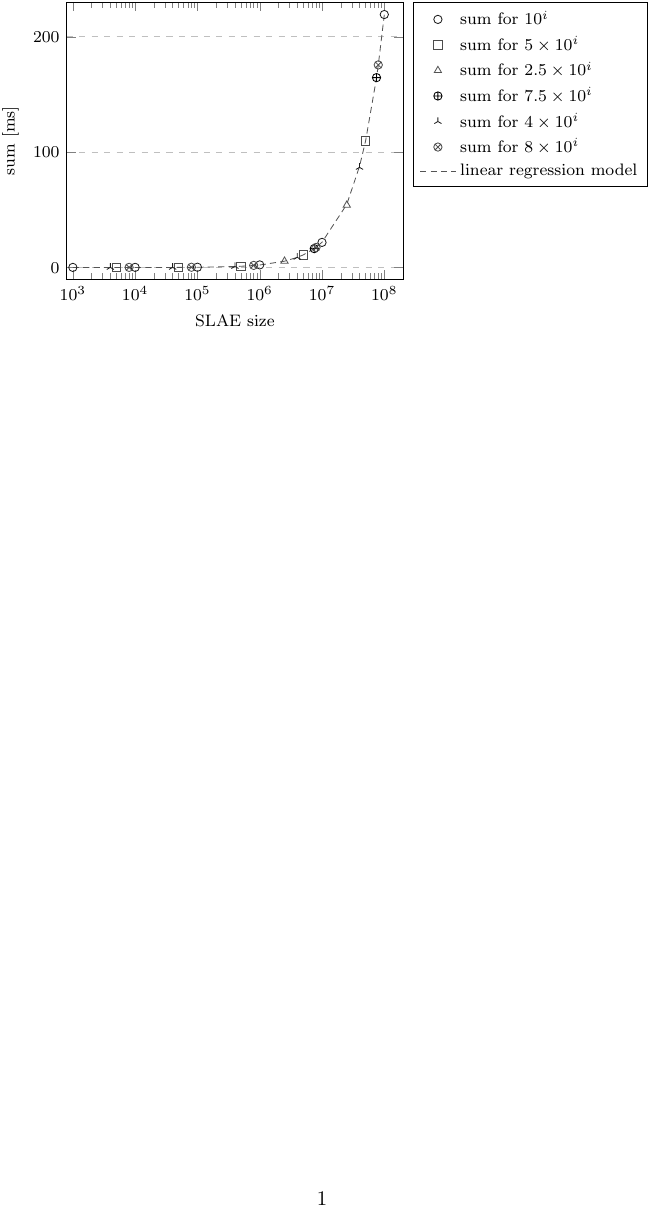}%{MMCP_aux_for_sum-crop.pdf}
\caption{Comparison of the sum of times for different SLAE sizes.}
\label{fig:sum_all_data_observed}
\end{figure}

%\label{sec:model_sum}
Let us define the sum of times of the GPU operations that take part in the stream overlap (see Eq.~\eqref{eq:time_comp}) as:
\begin{equation}
\textrm{sum} = T_{1}^{COMP} + T_{1}^{D2H} + T_{3}^{H2D} + T_{3}^{COMP}.
\end{equation}
According to the computational results (shown on Figure~\ref{fig:sum_all_data_observed}), $\textrm{sum}$ is linearly dependent on the SLAE size.
To build a heuristic for the optimum number of CUDA streams, we need to have a mechanism to predict the $\textrm{sum}$.
We perform regression analysis (supervised machine learning), where the dependent variable is the $\textrm{sum}$, and the independent variable is the SLAE size. The data for the $\textrm{sum}$ is split into training, and test set with the help of the \texttt{scikit-learn}~\cite{scikit} 
routine \texttt{train\_test\_split}, shuffle turned on, and splitting ratio $3:1$.
The obtained model (marked as {\it linear regressions model} on Figure~\ref{fig:sum_all_data_observed}) is:
\begin{equation}
\label{sum_model}
\textrm{sum\_model} = 0.0000021890017149\times\textrm{SLAE\_size}+ 0.1470644998564126.
\end{equation}
According to the R-squared coefficients (training: 0.9999813476643502, test: 0.9999942108504311) there is a perfect correlation between the observed and the predicted values. The mean squared error~(MSE) is 0.02 (test), the residual being smaller for smaller sums of times. The model is acceptable.
 
\begin{figure}[htb]
\centering
\includegraphics[trim={0cm 15cm 0 0}]{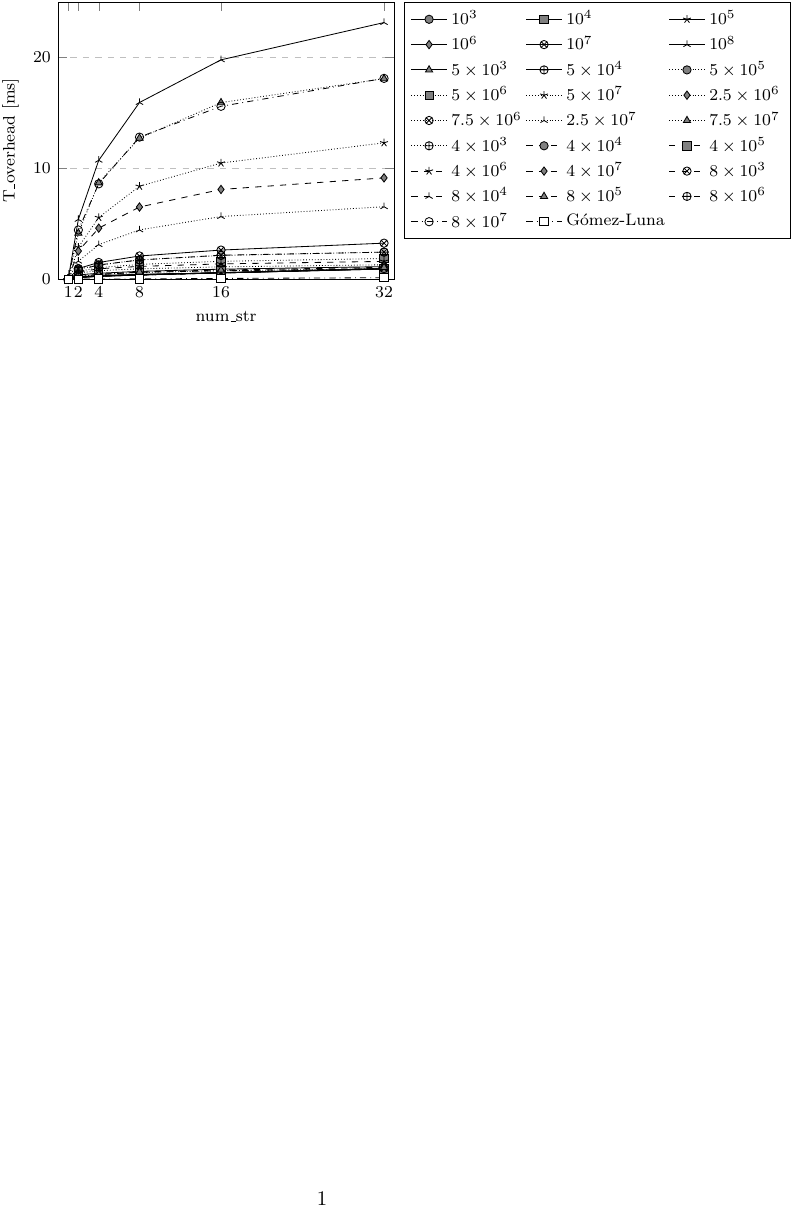}
\caption{$T\mathrm{\_{overhead}}$ for different SLAE sizes, on different number of CUDA streams.}
\label{fig:t_overhead_all_data}
\end{figure}

Let us define:
\begin{equation}
\label{eq:t_overhead}
T\mathrm{\_{overhead}} = (T_{str} - T_{non\_str}) + \frac{\textrm{num\_str}-1}{\textrm{num\_str}}\times \textrm{sum}.
\end{equation}
Figure~\ref{fig:t_overhead_all_data} shows that $T\mathrm{\_{overhead}}$ increases logarithmically with the increase in CUDA streams, following different patterns for SLAE sizes $\leq 10^{6}$, and $> 10^{6}$ (when the GPU starts getting utilized better). 
\begin{table}[htb]
%{\footnotesize
\centering
\caption{Times (given in [ms]) for the partition method for SLAE size $10^{6}$, for different number of streams.}
\label{tab:experimnets_10_6}
\begin{tabular}{|r|r|r|r|r|r|r|r|r|r|r|r|r|}
\hline
num\_str & $T_\mathrm{str}$ & $T_{\mathrm{non\_str}}$ & sum & $T{\mathrm{\_overhead}}$ & $\frac{\textrm{num\_str} - 1}{\textrm{num\_str}}\times$sum $-T{\mathrm{\_overhead}}$\\[0.5em]\hline
2 & 7.999136 & 8.817440 & 2.433568 & 0.398480 & 0.818304
\\\hline 
4 & 7.533248 & 8.817440 & 2.433568 & 0.540984 & 1.284192
\\\hline 
{\color{blue}8} & {\color{blue}7.401472} & 8.817440 & 2.433568 & 0.713404 & {\color{blue}1.415968}
\\\hline 
16 & 7.445952 & 8.817440 & 2.433568 & 0.909982 & 1.371488
\\\hline 
32 & 7.599968 & 8.817440 & 2.433568 & 1.140047 & 1.217472
\\\hline
\end{tabular}%}
\end{table}

\begin{figure}[htb]
\centering
\subfloat[]{{\fbox{\includegraphics[width=0.5\textwidth]{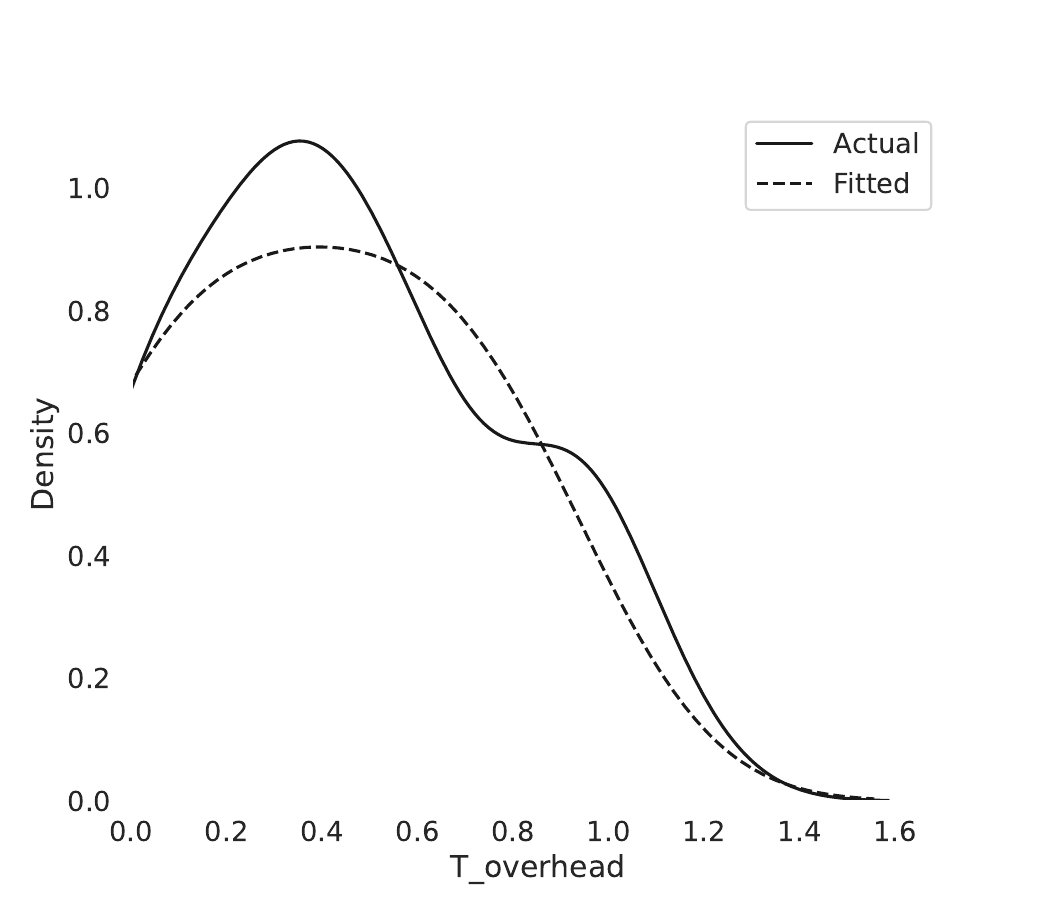}}}}\qquad
\subfloat[]{{\fbox{\includegraphics[width=0.5\textwidth]{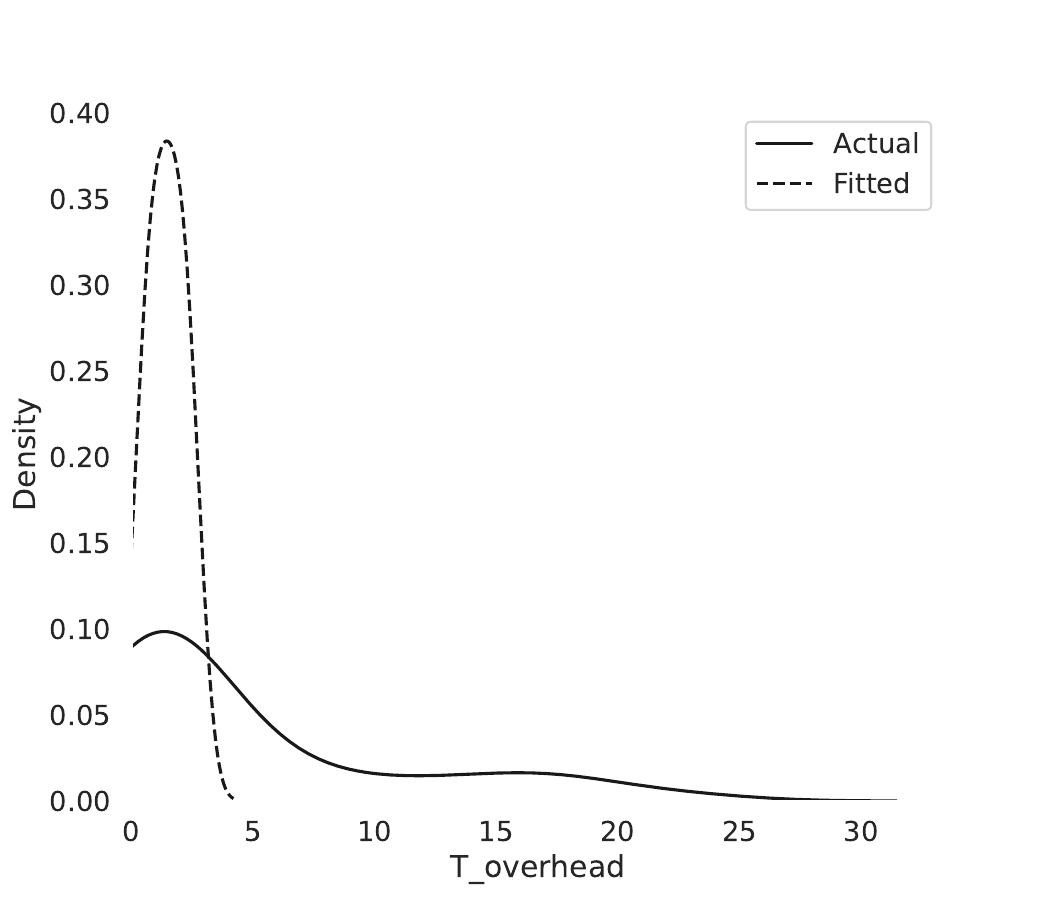}}}}%
\caption{Distribution of the actual vs.\,predicted values for the model of $T\mathrm{\_{overhead}}$ for SLAE sizes $\leq10^{6}$ (a), and $>10^{6}$ (b).}
\label{fig:dens_below}
\end{figure}

The optimum number of streams comes when (see Table~\ref{tab:experimnets_10_6} for an illustrative example):
\begin{equation}
\label{eq:opt_str}
T\mathrm{\_{overhead}} < \frac{\textrm{num\_str} - 1}{\textrm{num\_str}}\times\textrm{sum},
\end{equation}
%$T_{overhead}$ is smaller than the second term on the right-hand side of Eq.~\eqref{eq:t_overhead},
and
the difference between the term to the right-hand side of Eq.~\eqref{eq:opt_str} and $T\mathrm{\_{overhead}}$
is the biggest among the ones that fulfill the inequality.
According to the computational experiments, $T\mathrm{\_{overhead}}$ depends on both the SLAE size,
and the number of CUDA streams (Figure~\ref{fig:t_overhead_all_data}), and the trend is non-linear. 
The equations of the models (\textit{small} for SLAE sizes $\leq10^{6}$, and \textit{big} for SLAE sizes $>10^{6}$) are:
\begin{equation}
\label{t_ov_model}
\begin{aligned}
&\textrm{T\_overhead\_model\_small} 
= 0.0000002245645331\times\textrm{SLAE\_size}\\
&+ 0.6009426920043296\times\log_{10}(\textrm{num\_streams}) - 0.0605183610625299,\\[1em]
&\textrm{T\_overhead\_model\_big} 
= (0.0000000356594859\times\textrm{SLAE\_size}\\
&+ 0.0522781620855163)\times\log_{2}(\textrm{num\_streams}^{\frac{4}{3}}) + 0.3941472844770443.
\end{aligned}
\end{equation}
The two models are built with the help of the \texttt{SciPy}~\cite{scipy} 
routine \texttt{curve\_fit}, and the form of the functions is preset (different fitting curves were tested).
The data is split into training, and test set, shuffle turned on, and splitting ratio $3:1$. The dependent variable is $T\mathrm{\_{overhead}}$, and the independent variables are the SLAE size, and $\textrm{num\_streams}$.
The model metrics (Table~\ref{tab:t_overhead_model_metrics}) show that both the models are a good fit for the data.
\begin{table}[htb]
\caption{$T\mathrm{\_{overhead}}$ models' metrics.}
\label{tab:t_overhead_model_metrics}
\centering
%{\footnotesize
\begin{tabular}{|l|l|l|l|}
\hline
set & metric & model\_small & model\_big \\\hline
\multirow{3}{*}{training} & R-squared & 0.9531711290769591 & 0.9933780389080090\\
& MSE $=$ mean squared error & 0.0050126881205798 & 0.2451169015984794\\ 
& RMSE $=\sqrt{\textrm{MSE}}$ & 0.0708003398337877 & 0.4950928211946518\\\hline
\multirow{3}{*}{test} & R-squared & 0.9549695579010460 & 0.9896761975222511\\
& MSE $=$ mean squared error  & 0.0044441139999724 & 0.1447752928068124\\ 
& RMSE $=\sqrt{\textrm{MSE}}$ & 0.0666641882870588 & 0.3804934858927448\\\hline
\end{tabular}%}
\end{table}
Further, Figure~\ref{fig:dens_below} shows that the fitted values for SLAE sizes $\leq10^{6}$ are reasonably close to the actual values since the two distributions overlap almost completely, while the distribution of the fitted values for SLAE sizes $>10^{6}$ has a smaller variance compared to the actual ones.

\section{Discussion and Conclusions}
Using the algorithm described earlier, and the models for $\textrm{sum}$~(Eq.~\eqref{sum_model}), and $T\mathrm{\_{overhead}}$~(Eq.~\eqref{t_ov_model}), the optimum number of CUDA streams for different SLAE sizes are predicted (the results are summarized in Table~\ref{tab:opt_num_streams_observed_predicted}). 
For SLAE sizes $\leq10^{5}$ the optimum number of CUDA streams is 1, that is, the gain from using more streams is smaller than the overhead from creating them, while for SLAE sizes $\geq 4\times 10^{6}$ the optimum number is 32, because the SLAE size is big enough to require more GPU resources.
There are only two wrongly predicted values: for SLAE size $10^{5}$ the algorithm predicts 2 instead of 1 stream. The experiments show that the computational times for these number of streams are 1.239040~ms, and 1.230176~ms, respectively, so the difference is 0.008864~ms, which is negligible as a percentage of the computational time. The same is valid for size $5\times10^{5}$ (0.016224~ms difference between the computational times). Thus, the algorithm is acceptably good. The performance improvement we achieved is up to 1.30 (for SLAE sizes $8\times10^{7}$ and $10^{8}$).
The same ML-based approach could be applied to other applications as well.

\begin{table}[htb]
\caption{Optimum number of CUDA streams -- experiments ($N_{\mathrm{act}}$ vs.\,model predictions ($N_{\mathrm{pre}}$).}
\label{tab:opt_num_streams_observed_predicted}
\centering
%{\footnotesize
\begin{tabular}{|r|r|r|r|r|r|r|r|r|}
\hline
size & $N_{\mathrm{act}}$ & $N_{\mathrm{pre}}$ & size & $N_{\mathrm{act}}$ & $N_{\mathrm{pre}}$ & size & $N_{\mathrm{act}}$ & $N_{\mathrm{pre}}$\\\hline
$10^{3}$ & 1 & 1 & $4\times 10^{5}$ & 4 & 4 & $10^{7}$ & 32 & 32 \\\cline{4-6}
$4\times 10^{3}$ & 1 & 1 & $5\times 10^{5}$ & 8 & {\color{blue}\textbf{4}} & $2.5\times 10^{7}$ & 32 & 32 \\
$5\times 10^{3}$ & 1 & 1 & $8\times 10^{5}$ & 8 & 8 & $4\times 10^{7}$ & 32 & 32 \\
$8\times 10^{3}$ & 1 & 1 & $10^{6}$ & 8 & 8 & $5\times 10^{7}$ & 32 & 32 \\\cline{4-6}
$10^{4}$ & 1 & 1 & $2.5\times 10^{6}$ & 16 & 16 & $7.5\times 10^{7}$ & 32 & 32 \\\cline{4-6}
$4\times 10^{4}$ & 1 & 1 & $4\times 10^{6}$ & 32 & 32 & $8\times 10^{7}$ & 32 & 32 \\
$5\times 10^{4}$ & 1 & 1 & $5\times 10^{6}$ & 32 & 32 & $10^{8}$ & 32 & 32 \\\cline{7-9}
$8\times 10^{4}$ & 1 & 1 & $7.5\times 10^{6}$ & 32 & 32 & & &\\
$10^{5}$ & 1 & {\color{blue}\textbf{2}} & $8\times 10^{6}$ & 32 & 32 & & & \\\hline
\end{tabular}%}
\end{table}

\subsection{Experiments on other NVIDIA GPU cards.} 
The computational experiments that were performed on the basis of NVIDIA GPU card RTX A5000~\cite{a5000a,a5000b} showed that although this card has appox.\,1.25 times bigger peak memory bandwidth than NVIDIA RTX 2080Ti~\cite{2080a,2080b}, the stream heuristic preserves. The reason for this invariance most probably could be explained by the fact that the amounts of registers and shared memory on both the cards are the same, and these are the most utilized GPU resources by the algorithm.

\subsection{Experiments with FP32 precision.}
Some additional tests with FP32 precision were made.
The experimental results (summarized in Table~\ref{tab:opt_num_streams_fp32}) show that in 7 out of 16 SLAE sizes the optimum number of CUDA streams when using FP32 precision is half the optimum number of CUDA streams when using FP64 precision. This results were expected since when using single precision we need smaller number of memory transactions, and hence the memory utilization of the GPU is smaller than in comparison with the FP64 case. In the other 9 cases of SLAE sizes the optimum number of CUDA streams when using FP32 precision is the same as the optimum number of CUDA streams when using FP64 precision, and the difference between the computational times on these number of streams and half the number of streams is negligible as a percentage of the computational times. Overall, we recommend using the already built heuristic for FP64 precision, and to divide the optimum number of streams by two when using single precision.

\begin{table}[htb]
\centering
%{\footnotesize
\caption{The optimum number of CUDA streams for certain SLAE sizes according to the experiments with FP32 precision, and comparison with the optimum number of CUDA streams for FP64 precision.}
\label{tab:opt_num_streams_fp32}
\begin{tabular}{|r|r|r|r|}
\hline
SLAE size & optimum & optimum & comparison\\
& num\_str & num\_str &\\
& FP32 & FP64 & \\\hline
$\leq10^{5}$ & 1 & 1 & same\\\hline
$4\times 10^{5}$ & 2 & {\color{blue}\textbf{4}} & {\color{blue}\textbf{half}}\\\hline
$5\times 10^{5}$ & 4 & {\color{blue}\textbf{8}} & {\color{blue}\textbf{half}}\\
$8\times 10^{5}$ & 8 & 8 & same\\
$10^{6}$ & 4 & {\color{blue}\textbf{8}} & {\color{blue}\textbf{half}}\\\hline
$2.5\times 10^{6}$ & 16 & 16 & same\\
$4\times 10^{6}$ & 16 & {\color{blue}\textbf{32}} & {\color{blue}\textbf{half}}\\
$5\times 10^{6}$ & 16 & {\color{blue}\textbf{32}} & {\color{blue}\textbf{half}}\\\hline
$7.5\times 10^{6}$ & 32 & 32 & same\\
$8\times 10^{6}$ & 32 & 32 & same\\\hline
$10^{7}$ & 16 & {\color{blue}\textbf{32}} & {\color{blue}\textbf{half}}\\
$2.5\times 10^{7}$ & 16 & {\color{blue}\textbf{32}} & {\color{blue}\textbf{half}}\\\hline
$4\times 10^{7}$ & 32 & 32 & same\\
$5\times 10^{7}$ & 32 & 32 & same\\
$7.5\times 10^{7}$ & 32 & 32 & same\\
$8\times 10^{7}$ & 32 & 32 & same\\
$10^{8}$ & 32 & 32 & same\\\hline
\end{tabular}%}
\end{table}

\section*{Acknowledgement}
The authors would like to thank Dr.\,Alexander Ayriyan (JINR) for their
precious comments.

\IEEEpeerreviewmaketitle

\end{document}